\documentstyle[11pt,twocolumn,epsf]{article}
\begin{document}
\title{Determining Microscopic Viscoelasticity \\
in Flexible and Semiflexible Polymer Networks \\
from Thermal Fluctuations}
\author{B.~Schnurr, F.~Gittes, F.C.~MacKintosh, and C.F.~Schmidt \\
Department of Physics \& Biophysics Research Division, \\
University of Michigan, Ann Arbor, MI 48109-1120}
\date{July 28, 1997}
\maketitle

\begin{abstract}
We have developed a new technique to measure viscoelasticity in soft
materials such as polymer solutions, by monitoring thermal fluctuations
of embedded probe particles using laser interferometry in a microscope.
Interferometry allows us to obtain power spectra of fluctuating beads
from 0.1~Hz to 20~kHz, and with sub-nanometer spatial resolution. Using
linear response theory, we determined the frequency-dependent loss and
storage shear moduli up to frequencies on the order of a kHz. Our
technique measures local values of the viscoelastic response, without
actively straining the system, and is especially suited to soft
biopolymer networks. We studied semiflexible F-actin solutions and, as a
control, flexible polyacrylamide (PAAm) gels, the latter close to their
gelation threshold. With small particles, we could probe the transition
from macroscopic viscoelasticity to more complex microscopic dynamics.
In the macroscopic limit we find shear moduli at 0.1~Hz of
$G'=0.11\pm0.03$~Pa and $0.17\pm0.07$~Pa for 1 and 2~mg/ml actin
solutions, close to the onset of the elastic plateau, and scaling
behavior consistent with $G^*(\omega)\sim\omega^{3/4}$ at higher
frequencies. For polyacrylamide we measured plateau moduli of 2.0, 24,
100 and 280~Pa for crosslinked gels of 2, 2.5, 3 and 5\% concentration
(weight/volume) respectively, in agreement to within a factor of two
with values obtained from conventional rheology. We also found evidence
for scaling of $G^*(\omega)\sim\omega^{1/2}$, consistent with the
predictions of the Rouse model for flexible polymers.
\end{abstract}

\subsection*{Introduction}

The usefulness of synthetic polymeric materials, and the functions of
biopolymers, are all largely based on complex mechanical properties
that derive from a hierarchical structure.  The mechanical response of
polymer solutions or gels displays characteristics of both fluids
(viscosity) and of solids (elasticity), depending on the rate of change
of applied stress. This response is conventionally described by
frequency-dependent storage and loss shear moduli which are commonly
measured by active, mechanically-imposed oscillatory strain in
macroscopic samples, in contrast to stationary-flow geometries used for
measurements in fluids$^{1-5}$.

Such methods have also been used in the past to study reconstituted
biopolymer networks {\it in vitro}.  A general distinction between
synthetic polymers and biopolymers is that the former are typically
flexible, since their monomer size is small, whereas many biopolymers
are formed from large protein monomers and thus are much less flexible.
The viscoelastic properties of flexible and semiflexible polymer systems
are quite different. At a given volume fraction of polymer, the shear
modulus (or stiffness) of a semiflexible polymer network can be several
orders of magnitude larger than that of a flexible one. This may be one
reason why biological evolution has favored semiflexible polymer
networks for mechanical stability.  F-actin is one of the primary
components of the {\it cytoskeleton\/} of plant and animal cells, and is
largely responsible for the viscoelastic response of cells$^{6,7}$. The
$\sim$17~$\mu$m persistence length of actin filaments$^{8,9}$ is about
three orders of magnitude larger than their diameter. This tremendous
aspect ratio makes them ideal model semiflexible polymers. They are
rather rigid on the scale of cytoskeletal networks of typical cells,
which have characteristic mesh sizes on the order of microns.  In
reconstituted F-actin networks, individual filaments can be hundreds of
microns long, so that solutions of less than 0.1\% polymer (volume
fraction) are still strongly entangled.

Macroscopic rheological measurements in reconstituted F-actin solutions
have been reported in the literature$^{10-19}$. These experiments have
shown that the regime of linear viscoelastic response is small,$^{16}$
of order a few percent, and that this range decreases with increasing
F-actin concentration. For larger strains, strain hardening is observed,
followed by apparent shear thinning, which may be due to filament
breakage. This is to be expected for semiflexible polymer networks,
since the amplitudes of thermal undulations are small, and the full
extension of individual filaments is thus reached for small
strains,$^{20}$ making the response nonlinear. This makes measurements
with commercial rheometers problematic, although some custom intruments
can deal with strains substantially smaller than 1\%.$^{1}$ Active
macroscopic measurements on actin have given controversial
results$^{16}$ and nonlinearity may be one reason for this.

In attempts to approach {\it in vivo\/} conditions, viscoelasticity has
also been studied on a microscopic, cellular scale. Early work included
the manipulation of magnetic particles in gelatin $^{21}$ and in living
cells$^{22}$.  Recent experimental techniques for manipulating and
tracking sub-micrometer particles have revived the interest in {\em
microrheology}.  Several studies have used direct manipulation of
embedded magnetic beads in solutions of F-actin$^{23-27}$. These
experiments have been limited by the spatial and temporal resolution of
video microscopy.

\begin{figure}[h]
\epsfxsize=\columnwidth
\centerline{\epsfbox{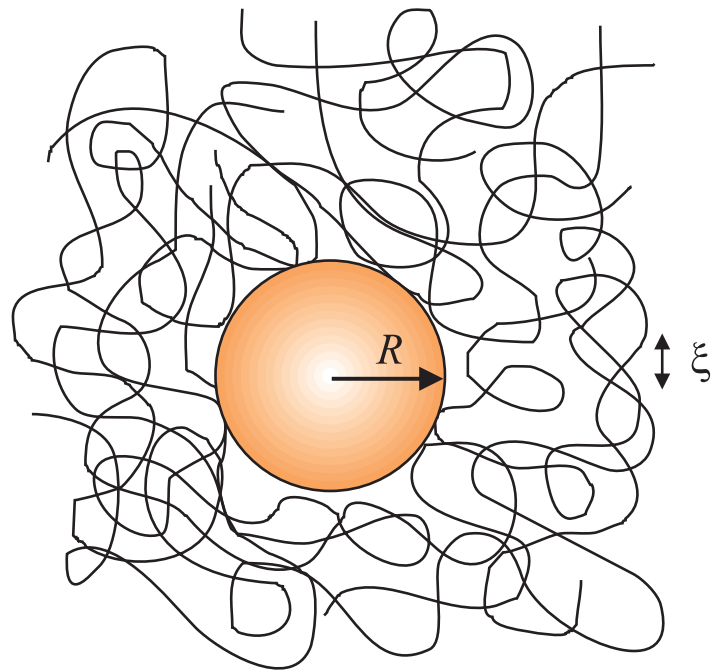}}
\caption[]{Schematic of a probe particle of radius $R$ embedded in an actin
network of mesh size $\xi$.}
\end{figure}

We introduce here a microscopic method of measuring viscoelastic
properties within micrometer-sized sample volumes.  We study entangled,
but non-crosslinked networks of semiflexible F-actin, as well as low
volume fraction gels of crosslinked polyacrylamide (PAAm), a flexible
polymer.  We observe the thermal fluctuations of micrometer-sized
particles embedded in soft gels (Fig.~1).  Laser interferometry in a
light microscope provides high resolution (less than 1~nm) and bandwidth
(from 0.1~Hz to 20~kHz). Using dispersion relations from linear response
theory, the frequency-dependent loss and storage shear moduli can be
determined from the fluctuation power spectra. The technique is in
principle less invasive than active methods, in that no strain at all is
imposed on the material.

Diffusing wave spectroscopy (DWS) has been used by others to observe
thermal fluctuations of ensembles of particles, and viscoelastic
properties in polymers and colloids have thereby been deduced$^{28,29}$.
This method measures average viscoelastic properties, as opposed to our
local measurements; we further compare this conceptually related method
with ours in the Discussion.

Our method measures shear moduli over a larger frequency range than
accessible to video based microrheology (while DWS and some other
macroscopic methods can reach higher frequencies$^{1,30}$).  Besides
providing a new way of passively measuring shear moduli in both
synthetic and biological polymer systems, our technique can be sensitive
to dynamics and material parameters that are inaccessible to macroscopic
methods: spatial inhomogeneities in networks can be studied on a
micrometer scale. At low frequencies, our method appears to be sensitive
to dynamics, such as non-shear deformations of the network with respect
to the solvent, that are not seen in macroscopic mechanical rheology.
For probe sizes comparable to the mesh size of the polymer network,
deviations from continuum elasticity become apparent. With decreasing
probe size, the transition from collective network dynamics to
single-filament dynamics can be traced to explore the microscopic basis
for macroscopic properties. This is particularly relevant for biopolymer
networks, with mesh sizes as large as microns. More importantly for
biology, microrheology will permit the characterization of small
samples, such as living cells.

\subsection*{Theory and Data Analysis Methods$^{31}$}

We model our experiments as embedding a spherical bead, of radius $R$,
in a linear viscoelastic medium of density $\rho$ with a macroscopic
complex shear modulus $G^*$ and Poisson ratio $\nu$.$^{32}$ For motions
of micron-sized probe particles, inertial effects are negligible at
frequencies $\omega$ where the inertial decay length (i.e., the inverse
magnitude $1/|\Gamma|$ of the complex propagation constant$^{1}$
$\Gamma=\left(\rho\omega^2/2G^*\right)^{1/2}$) is larger in magnitude
than $R$. At 1 kHz, $1/|\Gamma|$ is at least tens of microns in all our
samples. Correspondingly, inertial effects can be expected to be
relevant only for frequencies of order 1~MHz. In addition, the Reynolds
number for thermal motions on this scale is small: $\sim10^{-3}$.  In an
incompressible medium, bead motion is then determined entirely by the
frequency-dependent complex shear modulus
$G^*(\omega)=G'(\omega)+iG''(\omega)$, where $G'$ and $G''$ are the
storage and loss moduli.

In extremely low-frequency motions of a gel, the network deforms so
slowly that stress in the solvent relaxes completely, and the network
can undergo not only shear but compressional deformation, i.e. it
behaves as an elastic continuum with shear modulus $G\equiv
G^*(\omega=0)$ and Poisson ratio $\nu<1/2$. Stress $\sigma_{ij}$ is
related to strain $u_{ij}$ by$^{32}$
\begin{equation}
\sigma_{ij} = 2G \left[ u_{ij}+\nu\delta_{ij}
\Sigma u_{kk} / (1\!-\!2\nu) \right].
\end{equation}
The equation of elastic equilibrium, $\partial\sigma_{ij}/\partial x_i
=0$, can be solved exactly for $u_{ij}$ with no-slip boundary conditions
on a rigid spherical surface,$^{33-35}$ yielding an effective compliance
(reciprocal spring constant) for sphere displacement

\begin{equation}
\alpha(\omega=0)=\frac{1}{6\pi GR}\left[1+\frac{(\nu-1/2)}{2(\nu-1)}\right].
\end{equation}
The static compliance $\alpha$ in the limit $\omega\rightarrow 0$ is
significant because it is related by equipartition of energy to the
total mean square fluctuations of the probe particle in the gel:
\begin{equation}
\langle x^2\rangle=kT\alpha(\omega=0).
\end{equation}
Eqs.~(3) and (2) together provide an estimate of the static modulus $G$
from the variance $\langle x^2\rangle$ of fluctuations: a priori, $\nu$
is unknown in Eq.~(2) but over the relevant range, $0<\nu\le 1/2$, $\nu$
does not strongly affect the result; one can put $\alpha\approx 1/6\pi
GR$ with at most 25\% error. However, $\langle x^2\rangle$ is often
dominated by low-frequency instrumental drifts. We will instead use a
method (to be described below) of estimating $G^*(\omega=0)$ that is
less sensitive to such noise.

For frequencies above the $\omega\rightarrow 0$ limit, viscous stresses
develop in the gel, and we must reconsider Eq.~(2). The solvent and the
polymer network are coupled through viscous drag, which becomes stronger
with increasing frequency. Eventually, solvent and network are expected
to move as one at scales large compared with the network mesh size
$\xi$.  In this case the polymer network is strongly coupled to the
incompressible solvent and so behaves like an incompressible network,
which can be described by a Poisson ratio $\nu=1/2$. The crossover
frequency, above which incompressible behavior is observed, can be
estimated as follows. The viscous force per volume exerted by the
solvent on the network$^{36,37}$ is $\sim\eta v/\xi^2$, where $v$ is the
velocity of the solvent relative to the network. The local elastic force
per volume (exerted by the rest of the network) is $G \nabla^2 u \sim
u/R^2$ at the bead surface, where $u$ is the network displacement field.
Viscous coupling will dominate above a crossover frequency
\begin{equation}
\omega_c\simeq\frac{G}{\eta}\,\frac{\xi^2}{R^2}.
\end{equation}
Eq.~(4) is an order-of-magnitude estimate of $\omega_c$ that corresponds
to a few Hertz in our F-actin solutions ($G\simeq 1~{\rm Pa}$,
$\xi\simeq 0.1R$, $\eta=1~{\rm cP}$). Above $\omega_c$, provided that
$\xi$ is small compared with $R$, the bead moves as an inclusion in an
incompressible continuum viscoelastic medium. The stress-strain
relations in Eq.~(1) still hold, but with $\nu=1/2$. The spatial
equation of motion is still $\partial\sigma_{ij}/\partial x_i =0$ (at a
given $\omega$) and its solution is Eq.~(2) with $\nu=1/2$:
\begin{equation}
x_{\omega}=\alpha^*(\omega)f_{\omega}=\frac{1}{6\pi G^*(\omega) R}f_{\omega}.
\end{equation}
$\alpha^*(\omega)$ and $G^*(\omega)$ are complex because this formula
describes sinusoidal sphere displacements $x(t)=\alpha^*(\omega)f(t)$
caused by a force $f(t)\propto e^{-i\omega t}$. This is a
generalization of the well-known Stokes formula $f=6\pi\eta r \dot x$,
since $\eta=iG^*(\omega)/\omega$ in a purely viscous fluid.

We can rigorously justify Eq.~(5) directly, as follows.  It applies to
an incompressible medium without inertia.  The dynamics of the medium
are completely determined by its stress-strain relation (Eq.~(1)), which
becomes $\sigma_{ij} = 2G^*(\omega) u_{ij}$.  This is mathematically the
same law that describes a simple viscous fluid, but with the (complex
and frequency-dependent) quantity $iG^*(\omega)/\omega$ replacing the
viscosity $\eta$. Together with the equation of motion, $\partial
\sigma_{ij}/\partial x_i = 0$, the spatial equations to be solved are
also mathematically identical. The unique solution to these equations at
a frequency $\omega$, that satisfies no-slip boundary conditions on the
surface of the moving sphere, is precisely the Stokes law, with the
(complex and frequency-dependent) $G^*(\omega)$ replacing the quantity
$-i\omega\eta$. This has the simple and remarkable consequence that the
well-known Stokes force generalizes to give the correct compliance,
involving the complex $G^*(\omega)$.

For $\omega$ between zero and $\omega_c$, the solvent and network may
undergo significant relative motion. As a result of this partial
``draining'' of the network, Eq.~(5) will not be valid; the situation
must then be described by a ``two-fluid'' model$^{36,37}$.  Force
response and fluctuations may become dominated by long-wavelength modes
that relax over distances large compared with $\xi(G/\eta\omega)^{1/2}$.
This low-frequency draining allows for relative motion of network and
solvent, involving not only shear modes but also compressional modes in
the network. As the spatial extent of the modes becomes larger than the
bead size, the bead motion and its power spectrum are expected to become
independent of bead size.

The compliance $\alpha^*(\omega)=\alpha'(\omega)+i\alpha''(\omega)$ is the
complex response function for bead displacement. Thus, for frequencies
above $\omega_c$, Eq.~(5) and the fluctuation-dissipation theorem$^{38}$
relate $\langle x^2_\omega\rangle$, the power spectral density (PSD) of
thermal motion, to the imaginary part $\alpha''(\omega)$:
\begin{equation}
\langle x^2_\omega\rangle={4kT\alpha''(\omega)\over\omega}.
\end{equation}
Provided that $\alpha''(\omega)$ is known over a large enough frequency
range, one can recover the real compliance $\alpha'(\omega)$ from the
Kramers-Kronig relations$^{38}$ by evaluating a dispersion integral
\begin{eqnarray}
\alpha'(\omega)&=&
\frac{2}{\pi}P\int_0^\infty d\zeta\,\frac{\zeta\alpha''(\zeta)}{\zeta^2-
\omega^2}
\\
&=&\frac{2}{\pi}\int_0^\infty dt\,\cos\omega t\int_0^\infty
d\zeta\,\alpha''(\zeta)\sin(\zeta t).
\end{eqnarray}
This allows us to explicitly calculate (by Eqs.~(5) and (6)) the storage
and loss moduli $G'(\omega)$ and $G''(\omega)$. $P$ in Eq.~(7) denotes a
principal-value integral, meaning the $\epsilon\rightarrow 0$ limit of
the sum of two $\zeta$-integrals: from $0$ to $(\omega-\epsilon)$, and
from $(\omega+\epsilon)$ to $\infty$. However, as written in Eq.~(8),
this is equivalent to successive sine and cosine transforms of
$\alpha''(\omega)$, which are conveniently performed for the discrete
data points of a long time series. Standard Fourier routines$^{39}$ can
be used to evaluate the discrete PSD and then to perform the sine and
cosine transforms.  Either the PSD or the output of Eq.~(8) can be
smoothed (by averaging within bins of equal logarithmic spacing) without
strongly affecting the numerical results. As a control, we have also
calculated Eq.~(7) directly in the logarithmic domain, finding similar
results.

When the behavior of the network plus solvent is well described as a
single-component medium (so that Eq.~(5) holds), we can take the complex
reciprocal of $\alpha^*(\omega)$ to obtain the complex $G^*(\omega)$:
\begin{eqnarray}
G'(\omega)&=&\frac{1}{6\pi
R}\;\frac{\alpha'(\omega)}{\alpha'(\omega)^2+\alpha''(\omega)^2}\;,
\\
G''(\omega)&=&\frac{1}{6\pi
R}\;\frac{-\alpha''(\omega)}{\alpha'(\omega)^2+\alpha''(\omega)^2}\;.
\end{eqnarray}
Qualitatively, we expect the shear modulus $G^*(\omega)$ to exhibit a
characteristic form,$^{4}$ involving up to three distinct dynamical
regimes. If the polymer network is entangled but not crosslinked, as in
F-actin solutions, viscous flow will occur over times longer than the
reptation time $\tau_r$, with $G'(\omega)$ going to zero. The reptation
time is hours or days, for actin filaments tens of microns long,$^{40}$
and is thus not accessible to our experiments. For frequencies above
$1/\tau_r$, a rubber-like plateau appears, with a frequency-independent
elastic response $G'$, as in a crosslinked gel. Above the high-frequency
end of the plateau, the moduli are expected to increase with a
characteristic power of frequency, $G'$ and $G''\sim\omega^z$,
reflecting the increasingly limited relaxations of dynamic modes within
a mesh of the network$^{4}$.  Power law behavior is expected for the
shear moduli of any polymer network above the characteristic mesh
relaxation time and below the molecular high frequency cut-off, since
there is no other characteristic length or time scale in this regime. In
this scaling regime, from Eqs.~(5) and (6), the
PSD is then expected to follow
\begin{equation}
\langle x_\omega^2\rangle\sim\omega^{-(1+z)}.
\end{equation}

\subsection*{Materials and Experimental Methods}

Actin was purified ($\sim$95\% purity) from chicken skeletal muscle,
following standard recipes$^{41}$. Its concentration was determined both
by staining (BioRad) and by UV absorption at 290~nm (specific
absorption: 0.65~cm$^2$/mg). Monomeric actin (G-actin) was rapidly
frozen in liquid nitrogen and stored at $-85^\circ$C. Samples were
prepared by mixing G-actin with a small number of silica beads (Bangs
Laboratories, except for the 0.5~$\mu$m diameter beads, which were
kindly provided by E. Matijevic). After adding concentrated
polymerization buffer (F-buffer: 2~mM
hydroxyethyl-piperazineethanesulfonic acid (HEPES) (pH 7.2), 2~mM
MgCl$_2$, 50~mM KCl, 1~mM ethylenebis(oxyethylenenitrilo)tetraacetic
acid (EGTA), 1~mM adenosinetriphosphate (ATP)) the mixture was
immediately transferred into a sample chamber made from a microscope
slide, a cover slip and double-stick tape (with inner dimensions
15mm$\times$3mm$\times$70$\mu$m). In the sealed sample chamber the actin
polymerized at room temperature for at least one hour under slow
rotation. The samples were stored at 4$^\circ$C and examined within one
day (and in one case remeasured the following day, as a control).

Polyacrylamide (PAAm) gels were prepared according to a standard gel
electrophoresis recipe,$^{42}$ with concentrations of 2, 2.5, 3 and 5\%
(weight/volume) and a relative concentration of 3\% bis-acrylamide as a
crosslinker. Solutions were thoroughly degassed under vacuum, beads
added and polymerization initiated with tetramethylethylenediamine
(TEMED) and ammonium persulfate (APS). As with actin, polymerizing
solutions were then transferred into sample chambers
(15mm$\times$6mm$\times$140$\mu$m) and slowly rotated at room
temperature for at least one hour before starting experiments.

\begin{figure}[h]
\epsfxsize=\columnwidth
\centerline{\epsfbox{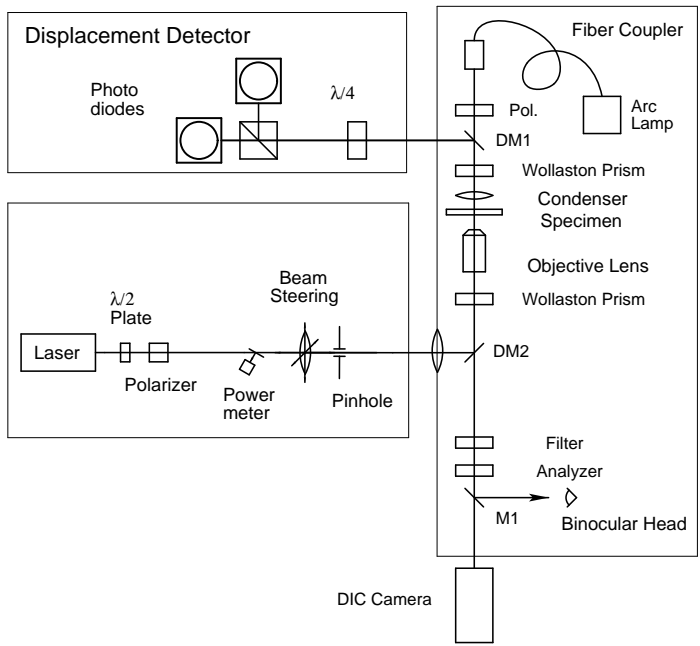}}
\caption[]{Schematic diagram of the custom-built inverted light microscope used
for differential interference contrast imaging and laser interferometric
displacement detection.}
\end{figure}

\begin{figure}[h]
\epsfxsize=\columnwidth
\centerline{\epsfbox{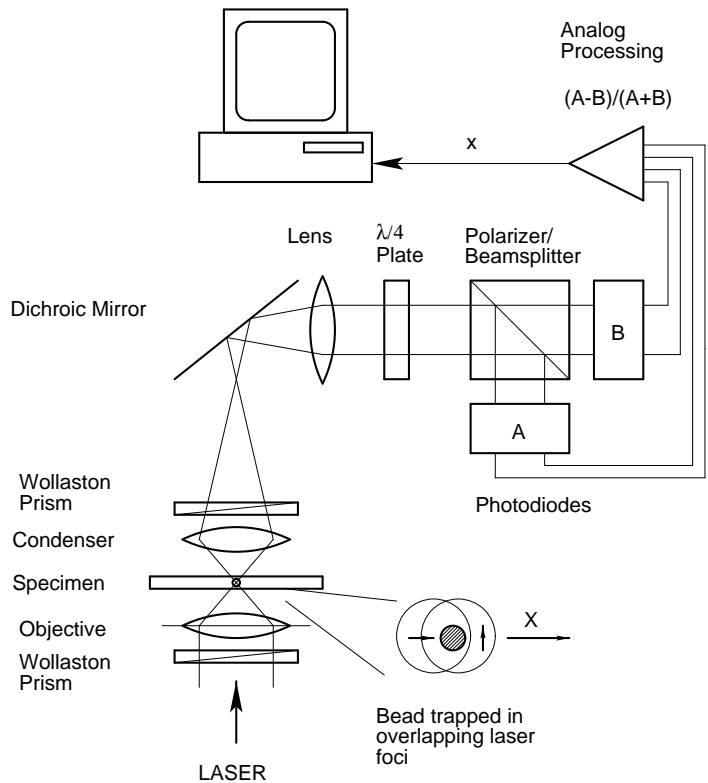}}
\caption[]{Schematic diagram of the interferometric displacement detection system.}
\end{figure}

Our microscope is a custom-built inverted instrument (optics from Carl
Zeiss, Inc.), constructed on an optical rail system and mounted on a
vibration isolated optical bench (Fig.~2). To detect the thermal motion
of beads imbedded in the gel we used an interferometer$^{43,44}$ (see
Fig.~3) with near-infrared laser illumination (1064~nm
NdVO$_4$, 3.4~W (cw) max.\ power, Topaz 106c, Spectra Physics).

A linearly polarized laser beam is split in two beams by the Wollaston
prism below the objective, which produces two diffraction limited
overlapping foci in the specimen plane. A refractive particle located
asymmetrically within the two foci will cause slightly elliptical
polarization after recombination of the two beams by the upper Wollaston
prism. A quarter-wave plate renders the light close to circularly
polarized, whose two perpendicular linear components are then detected
by two photodiodes. Deviations from circularity are calibrated to
measure particle displacements. The normalized difference between the
two signals is calculated by custom-built analog electronics (Fig.~3),
amplified and anti-alias filtered slightly below the Nyquist frequency.
This analog signal is then digitized and recorded using an A/D interface
(MIO 16X, National Instruments) and data acquisition software written in
Labview (National Instruments).

Although this is not done here, the focused laser beam can act, at high
enough laser power, as an optical trap, exerting forces on the particle
in the focus$^{44,45}$. Here, we want to measure unperturbed thermal
fluctuations, and thus need to minimize the trapping force. For that
purpose, a 1.5~mm diameter pinhole was inserted at a position conjugate
to the back focal plane of the objective, broadening the laser focus and
increasing the detector range. The laser power was typically 0.6~mW in
the specimen, low enough to make optical forces negligible.

Before recording fluctuations of an individual bead, it was centered in
the detector range using a piezo-actuated translation stage. For all
experiments reported here, the data acquisition rates were 50 or 60~kHz
(anti-alias filtered at 25~kHz). Time series data were recorded for at
least 17.5~seconds ($2^{20}$ data points) to obtain power spectra
ranging from 0.1~Hz to 25~kHz. Each spectrum was evaluated by Fast
Fourier Transform (FFT) after applying a Welch window. Power spectra
were smoothed by averaging within bins of geometrically increasing width
(with factor 1.1).

\begin{figure}
\epsfxsize=\columnwidth
\centerline{\epsfbox{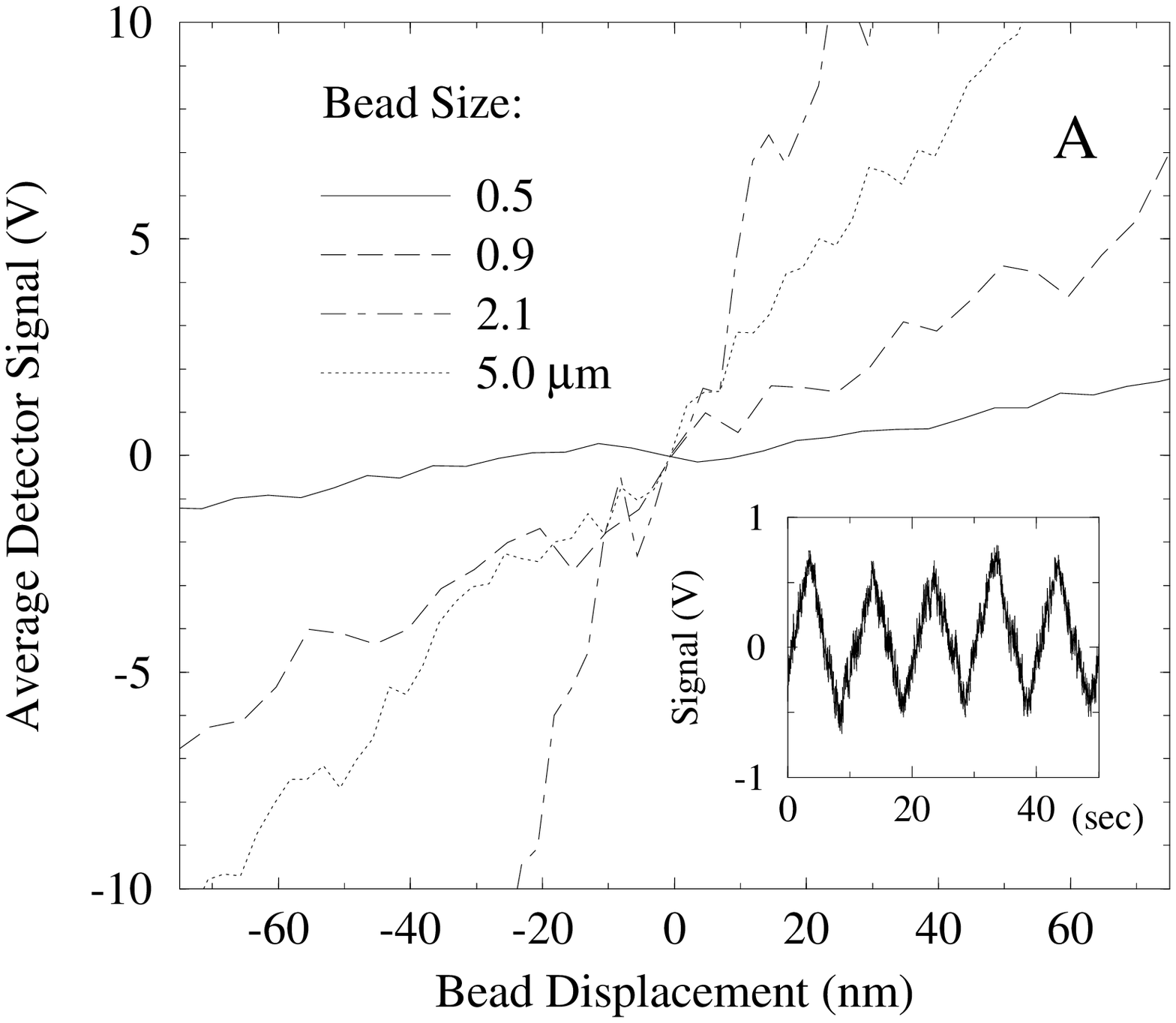}}
\epsfxsize=\columnwidth
\centerline{\epsfbox{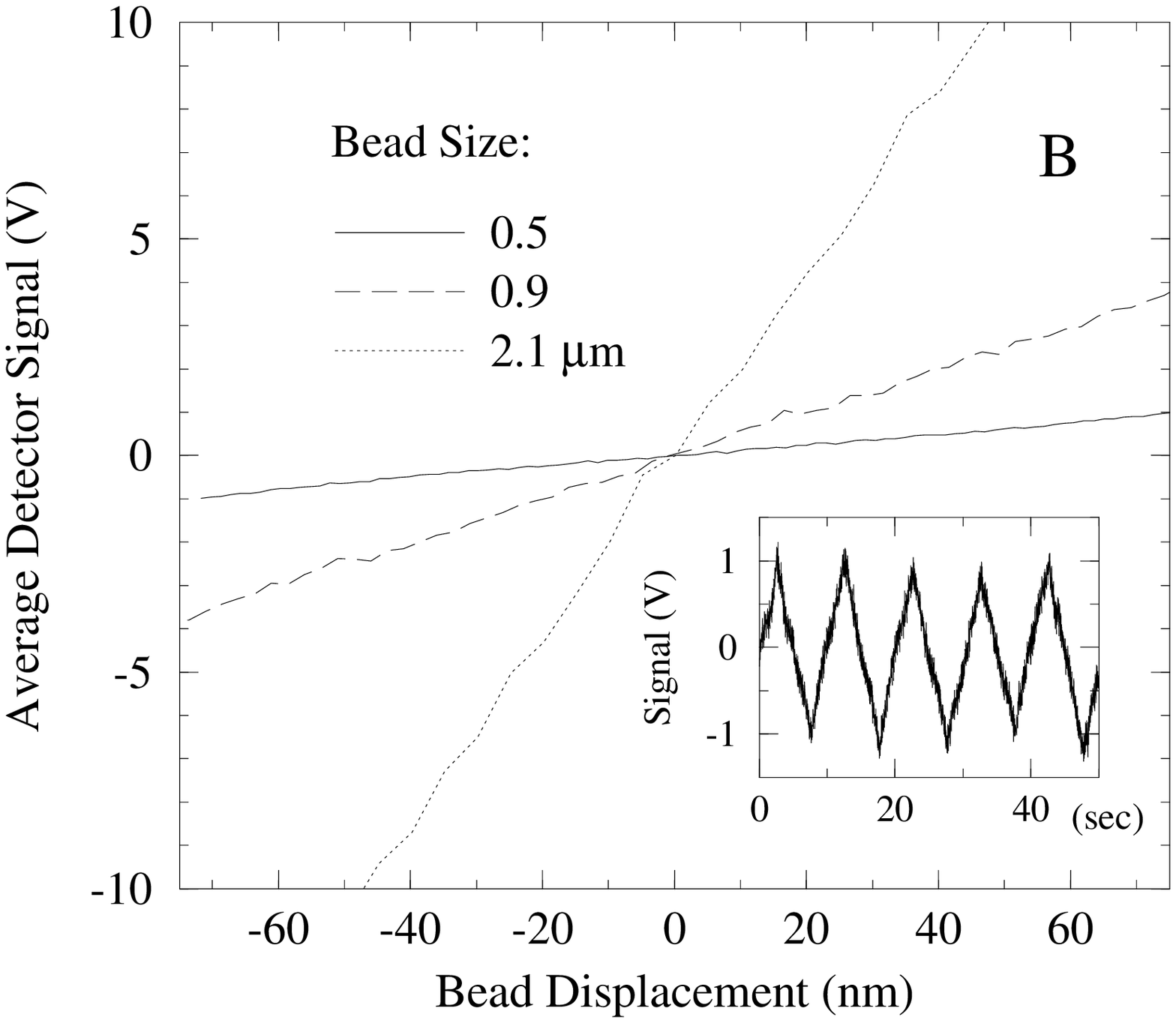}}
\caption[]{Detector sensitivity for displacement of silica beads of various
diameters embedded (A) in an actin solution (2~mg/ml F-actin) and (B) in
a 2.5\% PAAm gel, determined by driving a piezo-actuated stage with a
triangular signal at a frequency of 0.1~Hz. Response curves are averages
over several periods, sampled at 100~Hz (inset), smoothed (by a
Savitsky-Golay filter) and shifted vertically to pass through the
origin. For a uniform scale some curves include a gain factor. Inset:
Direct signal time series for a 0.5 $\mu$m bead.}
\end{figure}

The linear range of the detector is about 200~nm for 0.5~$\mu$m beads
and increases with bead size. Each observed bead was immediately
calibrated following the recording of thermal motion, by moving the
sample on the piezo-actuated stage with a triangular signal wave form,
producing constant velocity displacements (between 150~nm and 1~$\mu$m
peak-to-peak) at a frequency of 0.1~Hz. The driving voltage was
digitally synthesized to correct for the non-linearity and hysteresis of
the piezo actuators. Driver voltage (not shown) and detector response
(Fig.~4 inset) were recorded for about 40~seconds and
analyzed by plotting detector response versus bead displacement. Slopes
were estimated by linear regression from an averaged
response-displacement curve for each individual bead
(Fig.~4A and B). The result is a sensitivity factor which
is used to convert detector signals into actual displacements.

\begin{figure}
\epsfxsize=\columnwidth
\centerline{\epsfbox{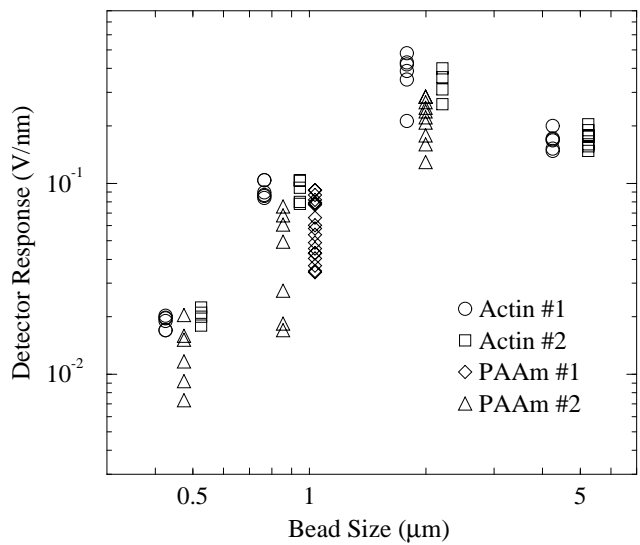}}
\caption[]{Bead-size dependence, and scatter, of detector response factors (see
Fig.~4). Four different experiments are compared; two with F-actin and
two with PAAm. Data points are horizontally shifted for clarity.
Vertical scatter is partly due to bead polydispersity, but PAAm
experiments show a distinctly stronger scatter. Bead diameters are 0.5,
0.9, 2.1, and 5.0~$\mu$m.}
\end{figure}

While centering the bead in the detector range, we found that the
sensitivity does not vary strongly in the perpendicular direction (Y),
out to Y-offsets comparable to the linear range.The sensitivity also did
not vary significantly with axial position in a range for which the bead
appeared focused in the video image ($\sim 0.6~\mu$m). Detector
sensitivities as a function of bead diameter are plotted in
Fig.~5. They increase roughly with the third power of
bead radius as long as the bead is smaller than the laser focus, since
the scattering amplitude is expected to scale with the volume of the
scatterer. For larger beads (in the limit of geometrical optics), the
sensitivity decreases roughly as $R^{-1}$, the only remaining length
scale. In other words, the same relative displacement $\Delta x/R$
should give the same signal $\Delta V$, independent of bead size:
$\Delta V\propto \Delta x/R$.

Investigating the origin of scatter in the detector sensitivity, we
found that variations in beam alignment dominated bead polydispersity
for the actin experiments. Bead polydispersity was checked directly
with transmission electron microscopy (TEM; data not shown) and found
to be about 10\%, with a non-normal distribution. The images showed both
larger beads, which appear to have formed by merging nuclei, and very
small beads that may be contamination. The scatter in the
displacement-response slopes was about $\pm$10\% (coefficient of
variation) for actin. When no correlation was evident between
calibration factors and the amplitudes of the PSD at around 100~Hz
(which should scale with bead radius $R$) we averaged the factor over
all beads observed in a sample.

For PAAm the scatter in sensitivities for a given bead size appeared
significantly larger ($\pm$30-60\% coefficient of variation), and
response factors were indeed correlated with the PSD, indicating that at
least part of the scatter was caused by bead size polydispersity. In
these cases, spectra were calibrated by the factors determined for
individual beads, which decreased the variance in the spectra between
beads in the same sample. Since the response factor scales roughly with
the third power of bead diameter for the smaller beads (see
Fig.~5), a 10\% coefficient of variance can explain a
30-40\% scatter in the spectra. The polymer network itself and possible
surface adsorption is not expected to influence apparent bead size and
thus sensitivity substantially. Furthermore, the index of refraction
changes little with polymer volume fraction in PAAm (measured to be
about a 1.2\% increase in relative index of refraction at 10\% polymer
w/v). Finally, surface chemistry that could produce a dense layer of
polymer directly on the bead surface is not expected to produce more
than a monolayer, and is thus limited to a thickness on the order of
tens of {\AA}.

\subsection*{Results}

\subsubsection*{Actin solutions}

Fluctuation time series were recorded for silica beads of 0.5, 0.9, 2.1,
and 5.0~$\mu$m diameter, embedded in actin gels of 1 and 2~mg/ml
concentrations. The fluctuation signal is well above the noise, as shown
by comparison to spectra of a bead immobilized on a surface, and of the
water background without bead (Fig.~6). These controls display
the effects of low-frequency instrumental noise due to thermal drifts in
the microscope and the laser beam path, acoustic vibrations, and
beam-pointing fluctuations. At high frequencies, noise depends on laser
intensity since detection shot noise and preamplifier noise are dominant
there. We emphasize that the amplitude of this noise is small, and
remains about two orders of magnitude below the signal. Note that all
spectra are (analog) anti-alias filtered at 25~kHz, causing the
``tails'' at about 20~kHz.

\begin{figure}
\epsfxsize=\columnwidth
\centerline{\epsfbox{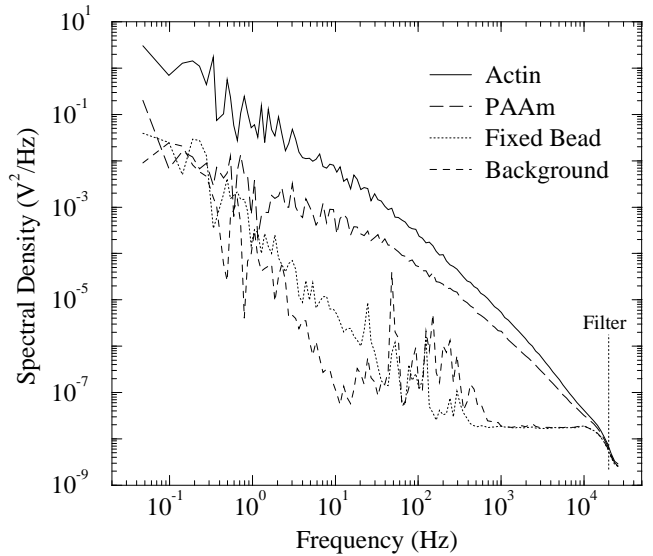}}
\caption[]{PSDs (power spectral densities) of background noise compared to
signal for F-actin and PAAm. Voltage signals were sampled at 50~kHz
and spectra were smoothed by averaging in equal log bins. An
anti-aliasing filter attenuates the signal strongly above about
20~kHz. The lower two curves show the PSD of the interferometer signal
for water without bead, and with a 0.5~$\mu$m bead fixed on a glass
surface. The F-actin spectrum (0.5~$\mu$m bead in 2~mg/ml actin) is
everywhere about two orders of magnitude above the noise, whereas the
PAAm spectrum (0.5~$\mu$m beads in 2\% PAAm) is dominated by noise
below about 1~Hz. }
\end{figure}

The power spectra of bead fluctuations in actin solutions were
reproducible in shape and amplitude between different beads in the
same sample (Fig.~7A), as well as between samples
(Fig.~7B). Observed coefficients of variance (of the PSD
at 100 Hz in Fig.~7A) within a sample (N=6) were 22\%
(0.5~$\mu$m), 24\% (0.9~$\mu$m), 8.7\% (2.1~$\mu$m) and 15\% (5.0~$\mu$m
bead diameter). Fig.~7B shows two spectra from samples
in different experiments but from the same protein preparation, and one
made with actin from a different laboratory (sample provided by Paul
Janmey). We find that beads in different samples do not scatter more
than different beads within one sample.

Slow changes have been reported to occur in actin solutions over many
hours after polymerization, possibly due to the depletion of ATP in the
buffer, or to the annealing of filaments. To address such effects, we
remeasured samples the day following their preparation, without seeing
significant differences in the spectra (Fig.~7C).
Observed beads were typically located at heights between 10 and
50~$\mu$m above the sample chamber surface. Variations in the spectra
were not correlated with distance from the boundaries (data not shown).

\begin{figure}
\epsfxsize=\columnwidth
\centerline{\epsfbox{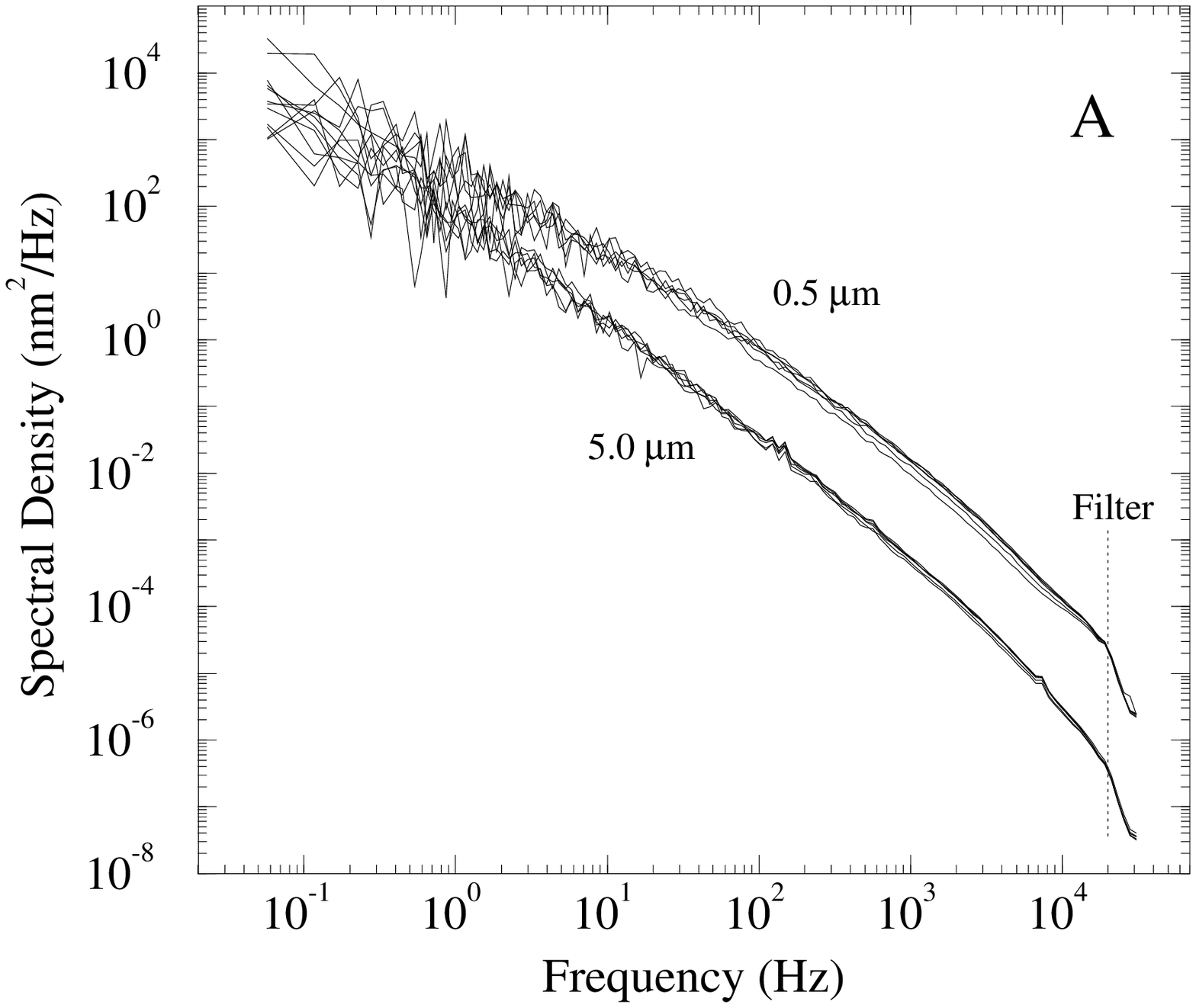}}
\epsfxsize=\columnwidth
\centerline{\epsfbox{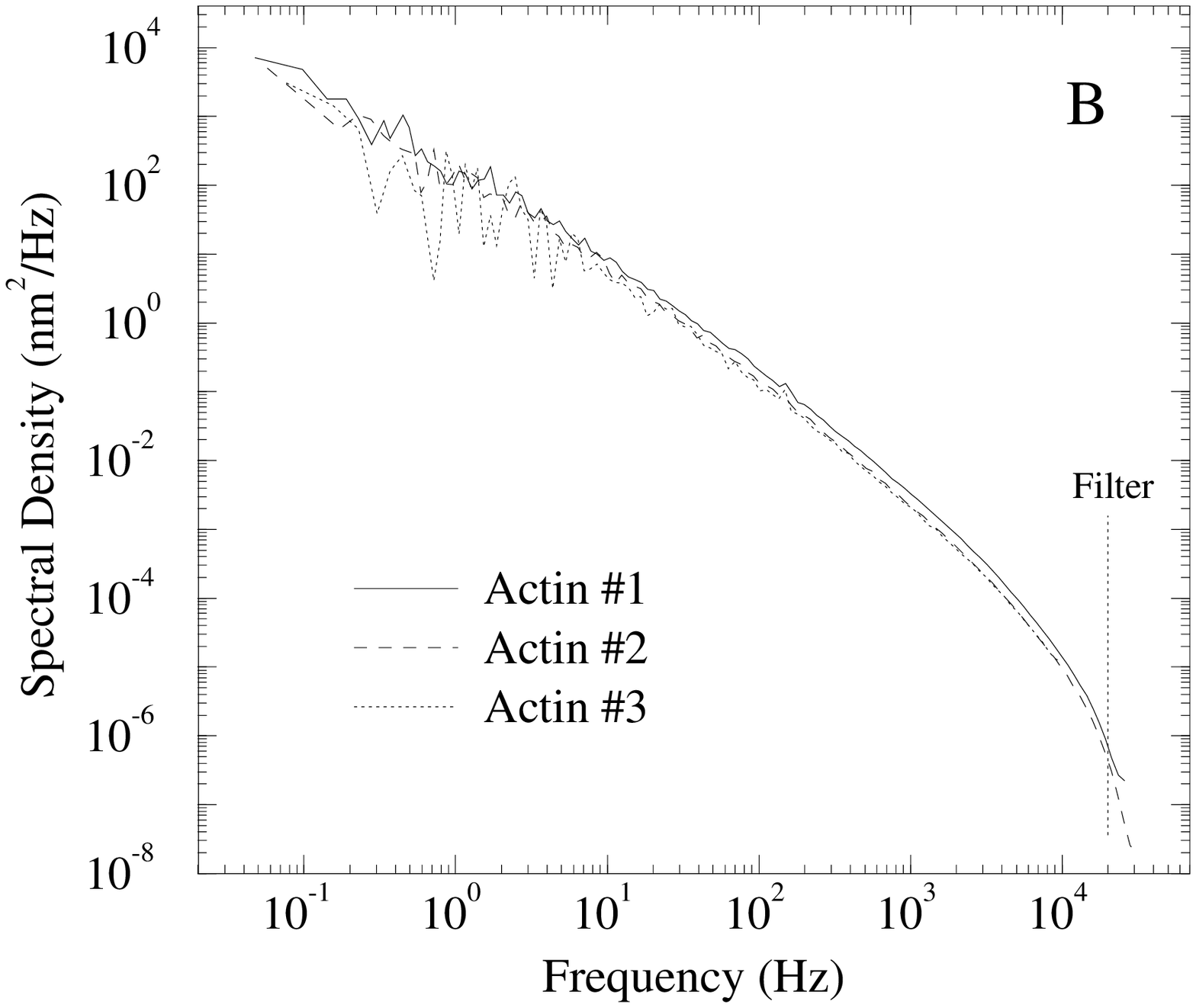}}
\end{figure}

\begin{figure}
\epsfxsize=\columnwidth
\centerline{\epsfbox{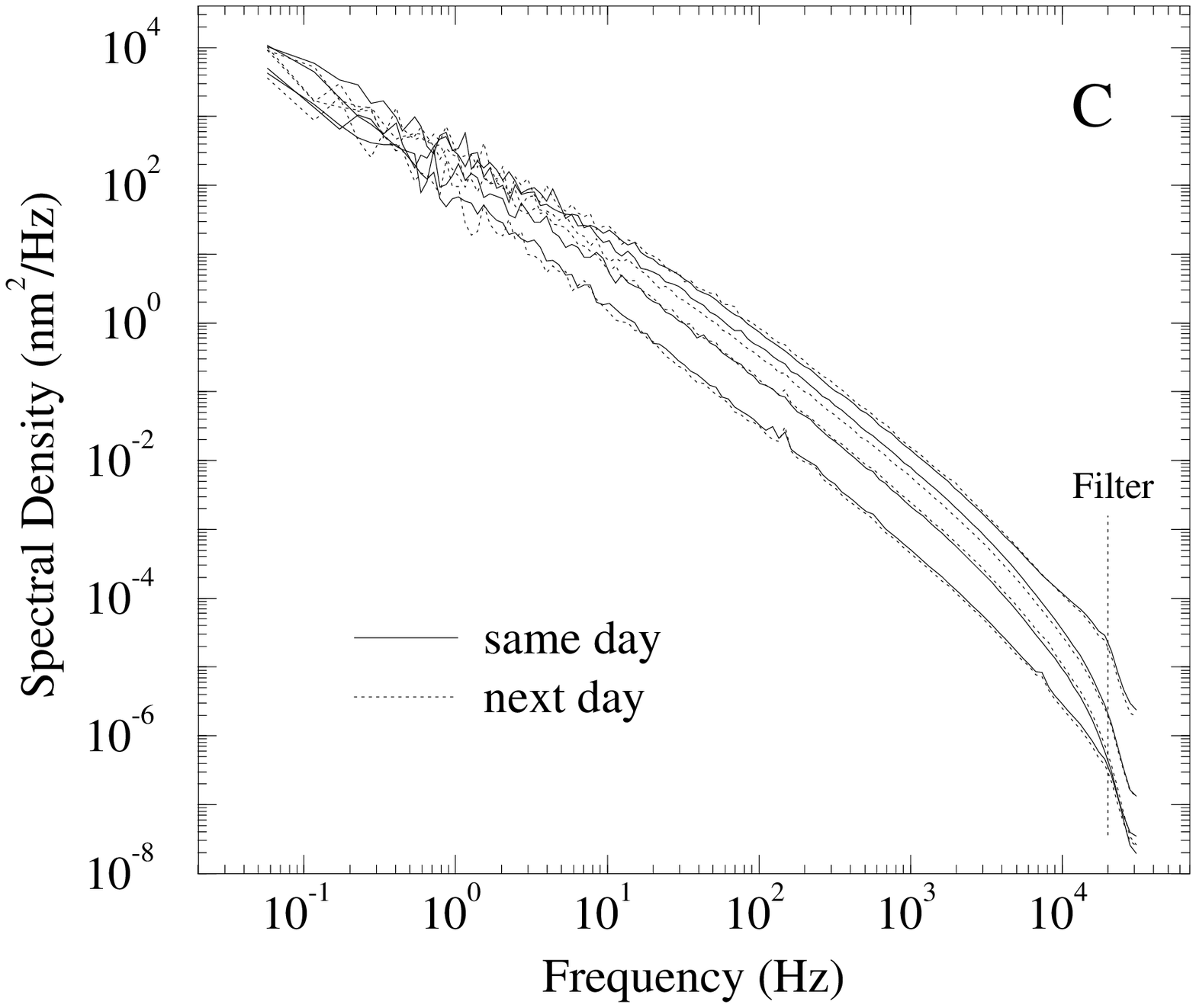}}
\epsfxsize=\columnwidth
\centerline{\epsfbox{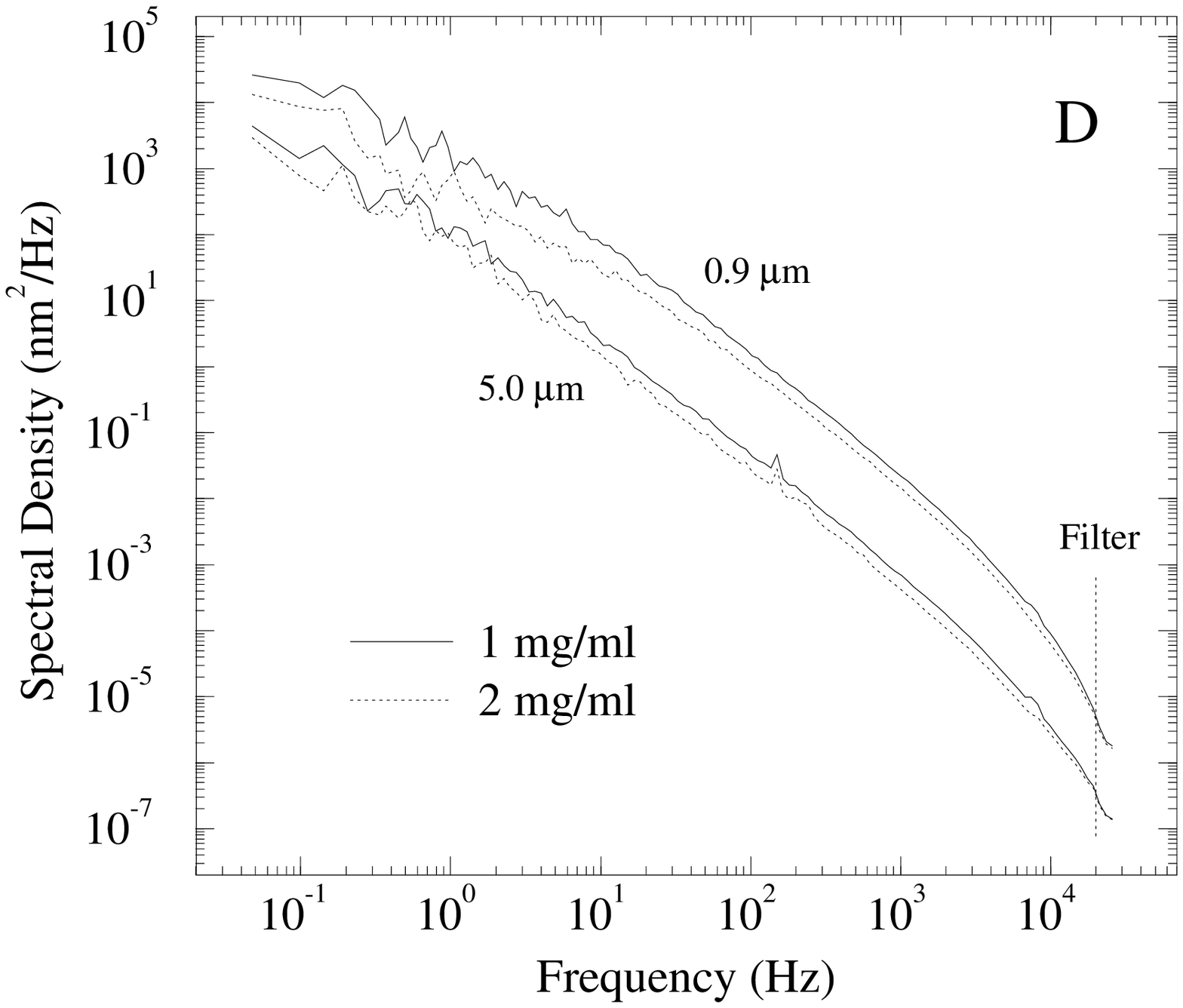}}
\caption[]{(A) Reproducibility within individual samples of F-actin (2 mg/ml).
Six beads each with diameter 0.5 and 5.0~$\mu$m. Height above the sample
cell surface was between 10 and 50~$\mu$m. (B) Reproducibility between
experiments and different protein preparations. Spectra from two
experiments using the same actin preparation at a concentration of
2~mg/ml (\#1: 1.8~$\mu$m, \#2: 2.1 $\mu$m silica beads) and using actin
from a different laboratory (P.~Janmey) (\#3: 1.8~$\mu$m silica beads).
(C) Aging of F-actin samples. Spectra from four samples (0.5, 0.9, 2.1,
5.0 $\mu$m beads, top to bottom, with six beads averaged per sample)
measured within hours of polymerization and remeasured the following
day.  (D) Comparison of 1 and 2~mg/ml F-actin solutions. Spectra from
two samples with 0.9~$\mu$m beads and two samples with 5.0~$\mu$m beads.
Six beads are averaged in each plotted PSD.}
\end{figure}

Power spectral densities (PSDs) of thermal motions of 0.9 and
5.0~$\mu$m silica beads in actin solutions at different concentrations
are shown in Fig.~7D. The shape of 1 and 2~mg/ml
spectra is similar and the PSD amplitude of the less concentrated
sample is smaller, as expected.

A phase transition to a nematic phase has been reported for actin at
concentrations of about 2~mg/ml$^{40,46}$.  However, we see no
qualitative difference in the appearance or in the behavior of our actin
gels at 1 and 2~mg/ml. There were no signs of inhomogeneities under
polarization microscopy.  Furthermore, we do not see any greatly
increased scatter in our measured values of $G^*$, which would be
expected in a highly anisotropic medium, given our uniaxial detection
technique.

\begin{figure}
\epsfxsize=\columnwidth
\centerline{\epsfbox{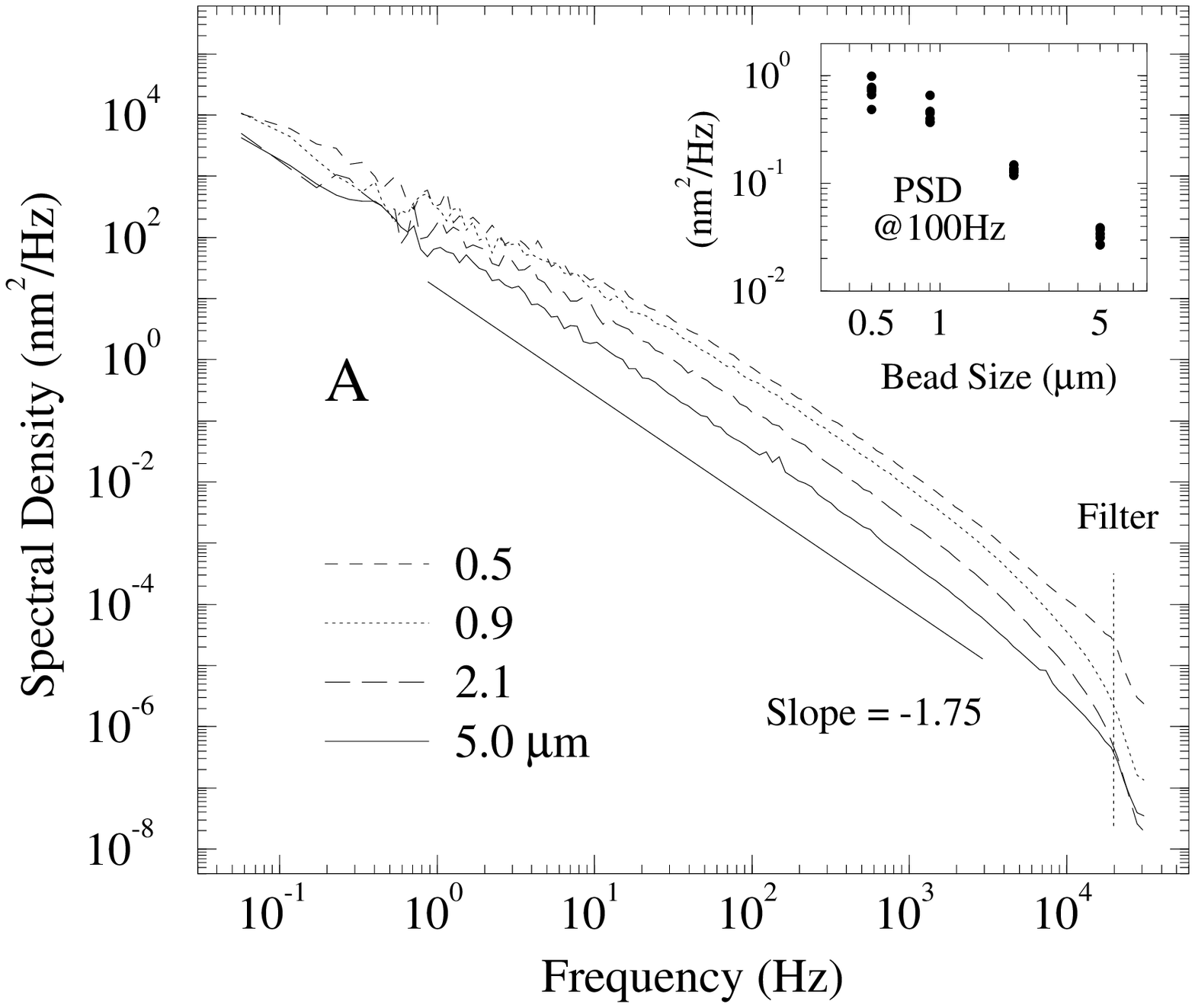}}
\epsfxsize=\columnwidth
\centerline{\epsfbox{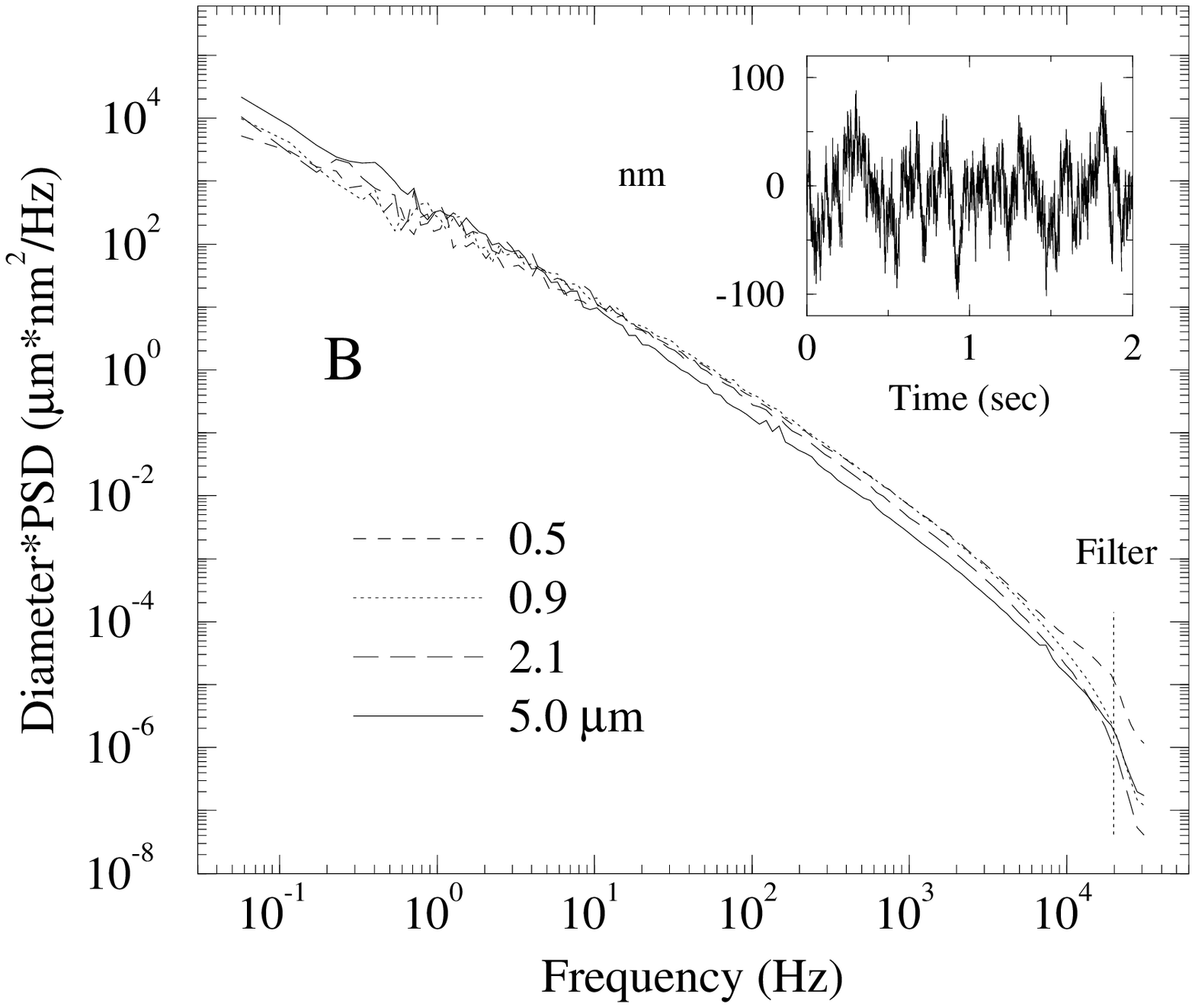}}
\caption[]{(A) Bead-size dependence of PSDs in 2~mg/ml F-actin solutions using a
60~kHz sampling rate. Spectra were calibrated with response factors
found as in Fig.~4. PSDs from beads between 0.5 and 5.0~$\mu$m diameter
are shown. For each bead size PSDs from six different beads in one
sample were averaged. The line indicates a power-law slope of $-1.75$.
Inset: PSD amplitudes at 100~Hz of individual smoothed spectra that make
up the averages. (B) Averaged PSDs from (A) multiplied by their
respective bead diameters. Inset: direct time series data for 0.9~$\mu$m
bead.}
\end{figure}

Fig.~8A shows averaged and smoothed spectra obtained from
2~mg/ml actin solutions with beads ranging from 0.5 to 5.0~$\mu$m in
diameter. The variance in the smoothed spectra (values at 100~Hz) as a
function of bead diameter is plotted in the inset. Part of the scatter
presumably reflects local inhomogeneities in the networks. Smaller
beads, which probe more locally, should show more scatter. For
0.5~$\mu$m beads, however, some variation is also due to drift out of
the linear range of the detector, and to bead polydispersity, as
discussed in Materials and Experimental Methods.  Detector sensitivity
depends on bead size, as we have shown in Fig.~5. The
high frequency noise level follows this dependence, with the smallest
(0.5~$\mu$m) and largest (5.0~$\mu$m) beads showing the largest relative
noise levels.

The largest beads are about a factor of 10 to 20 larger than the average
mesh size (0.25~$\mu$m for actin at 2~mg/ml$^{47}$) and a continuum
elastic model should apply.  Approximate power law behavior, with a
slope of about $-1.75$, is evident over about three decades in frequency
for the 5~$\mu$m beads, as indicated in Fig.~8. All spectra show a
slight downturn at about 3~kHz, an effect which we cannot yet explain.
At the high frequency end, the data have not reached the noise bottom
evident in the spectrum of a bead fixed to the substrate surface
(Fig.~6).

Eq.~(5) predicts that spectra should scale with the bead radius $R$ for
high frequencies. This is roughly observed as shown in Fig.~8B, but a
systematic deviation from $R$-scaling is evident. Small beads show power
spectral densities of higher amplitude than expected. We believe that
the increased fluctuations are not due to diffusion through the network
or constraint release due to filament reptation; in video recordings of
up to two hours we did not observe any long-range diffusive motion.
Furthermore, we can exclude that the beads themselves bind to F-actin
(with potential bunching of filaments as a result), because we observed
practically unimpeded diffusion of silica beads (0.2~$\mu$m) with
diameters below the mesh size.

We instead attribute the lack of scaling to the steric exclusion of
filaments near the bead. This model is elaborated quantitatively in a
separate publication$^{34}$ but given qualitatively in the following. A
large bead of radius $R$ will affect the distribution of filaments to an
exclusion depth $\ell\sim\ell_p$, where $\ell_p$ is the filament
persistence length. Since the Stokes flow field around a large bead
extends beyond the exclusion depth to a distance on the order of $R$,
large beads will mainly see the unperturbed network according to
Eq.~(5). Small beads ($R\ll \ell_p$) will exclude filaments to a depth
of the order of $R$, and thus see an effective compliance larger than
that described by Eq.~(5). Once in this limit, the compliance should
again scale as $R^{-1}$. This is consistent with our observations: the
spectra for the smaller beads (0.5 and 0.9~$\mu$m) do indeed superimpose
after scaling (Fig.~8B).

At low frequencies the slopes of the spectra decrease, and the PSDs
converge for all beads (Fig.~8A). The onset of the slope change,
however, {\em does\/} depend on bead size. Therefore, for the actin
samples it can {\em not\/} be the onset of the elastic plateau as is the
case for the PAAm data shown below. Instead, we interpret the change in
slope as the transition, predicted by Eq.~(4), from a regime of pure
shear fluctuations to long-wavelength draining modes, for which the
spectra should become bead-size independent$^{34}$. Such a transition is
not seen in PAAm because the smaller mesh size inhibits draining.  The
estimate for the crossover frequency for our F-actin solutions---on the
order of 10~Hz---is also consistent with the data.

Fig.~9A-D show the storage and loss shear moduli $G'(\omega)$ and
$G''(\omega)$ obtained for various bead sizes in 2~mg/ml actin solutions
by evaluating the Kramers-Kronig integral, Eq.~(7). The continuum model
should apply to the largest beads (Fig.~9D) which are about 20 times
larger than the mesh size. There, scaling is observed directly in both
$G'$ and $G''$ in the range from about 5 to 100~Hz. Pure scaling
corresponds theoretically (via Eqs.~(5) and (7)) to a complex modulus
$G^*(\omega)\propto (i\omega)^{z}$, so that their ratio is $G''/G'=\tan
\pi z/2$$^{4}$. This ratio, averaged between 10 and 100~Hz for the
largest beads (Fig.~9D), allows a precise estimate of $z=0.75\pm0.02$.

\begin{figure}
\epsfxsize=\columnwidth
\centerline{\epsfbox{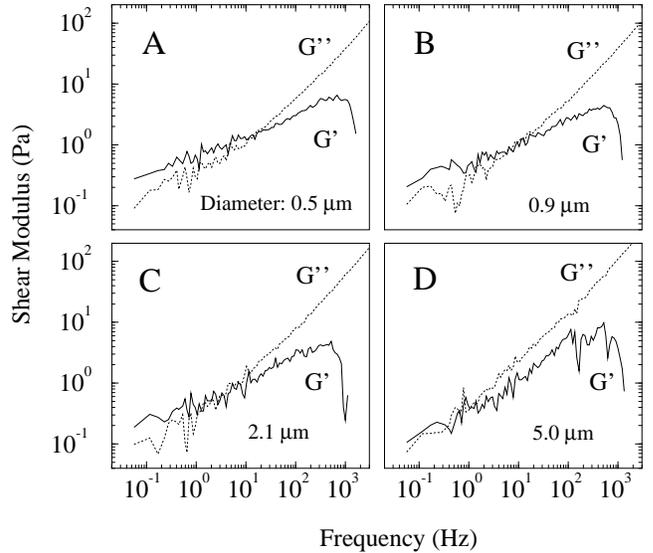}}
\caption[]{(A-D) Real and imaginary parts of the complex shear modulus of 2~mg/ml
actin solutions, obtained from the spectra in Fig.~8 via the
Kramers-Kronig relation (Eq.~(7)) for bead diameters 0.5, 0.9, 2.1 and
5.0~$\mu$m. Power law scaling of the complex shear modulus emerges for
the largest beads (D). At low frequencies, an elastic plateau is not yet
apparent.}
\end{figure}

Eqs.~(5) and (6) imply scaling of the power spectrum, via Eq.~(11), as
$\langle x^2_\omega\rangle\propto\omega^{-1.75}$, consistent with the
apparent slope in (Fig.~8A). The scaling exponent $z$ of about 3/4 in
the macroscopic behavior is distinct from the predictions of Rouse-like
scaling with exponent 1/2 observed for flexible polymers (see PAAm
below). We are unaware of any existing model that predicts $z=3/4$ for
the scaling of the complex, macroscopic shear modulus.

We also note that no plateau is (yet) visible in $G'(\omega)$. To
compare with the literature, we quote values at the lowest frequencies
we observed. $G'$ at 0.1~Hz was $0.11\pm0.03$~Pa for 1~mg/ml and
$0.17\pm0.07$~Pa for 2~mg/ml (actin). Ruddies {\it et al.\/}$^{15}$
measured a value of $G'\approx 0.3~Pa$ near the observed onset of the
plateau in 0.3 mg/ml actin.

For smaller beads, the apparent $G'(\omega)$ and $G''(\omega)$,
calculated under the assumptions of a single fluid behavior, show
deviations from macroscopic shear elastic behavior (Fig.~9ABC), and a
pure scaling regime is not reached in the data. This confirms that the
assumption of an incompressible network is not valid over the whole
frequency-range of the spectra for the smaller beads.

\subsubsection*{Polyacrylamide gels}

Polyacrylamide gels, even at volume fractions close to the gelation
threshold, are so rigid that we approach the limits of our technique.
Given the noise floor illustrated in Fig.~6, our technique is limited at
present to static shear moduli of up to a few hundred Pascal.  A volume
fraction of 2\% (with relative crosslinker concentration of 3\%) was the
lowest concentration that resulted in reproducibly solidified gels.
Fig.~10 shows PSDs of 0.9~$\mu$m beads embedded in 2, 2.5, 3, and 5\%
(volume fraction) PAAm gels, linearly averaged over several beads
respectively, and log-binned as described in Materials and Experimental
Methods. Absence of long-range diffusion of the beads confirms that all
gels were crosslinked into solids. Since the mesh size in the PAAm gels
(about 5~nm in the 2\% gels$^{48}$) is in all cases much smaller than
the bead size, a continuum elastic model is expected to apply
throughout.

\begin{figure}
\epsfxsize=\columnwidth
\centerline{\epsfbox{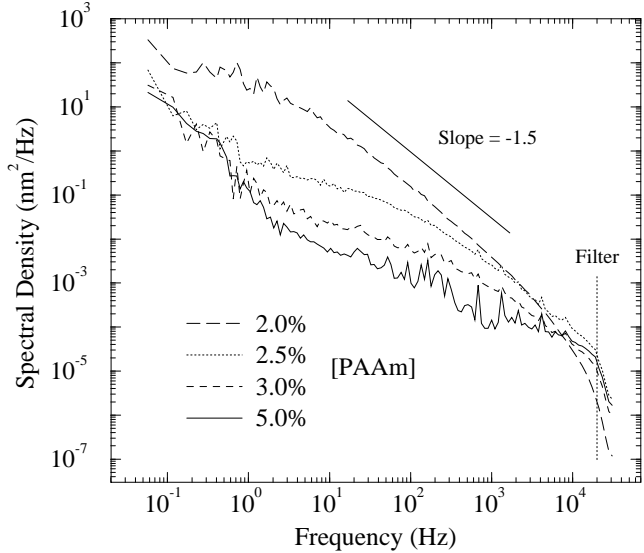}}
\caption[]{Concentration dependence of PSDs for 0.9~$\mu$m beads in PAAm gels. The
PSD from the 5\% gel is substantially affected by instrumental noise.
The line indicates a power law slope of $-1.5$. The onset of the elastic
plateau is visible as a concentration-dependent change in slope at low
frequency.}
\end{figure}

PSDs for beads in the softest gels (2\%) show a power law slope of about
$-1.5$ near 100~Hz, which is distinctly less steep than in F-actin. We
again observe the slight but unexpected downturn in the spectra above
about 3~kHz. The flattening of the spectra below a few Hertz reflects
the elastic plateau (compare Fig.~11A). This plateau extends higher for
stiffer gels. For concentrations above 2.5\%, instrumental noise begins
to dominate the low-frequency part of the spectra (and approaches the
fixed-bead spectrum, see Fig.~6). Raising the PAAm concentration higher
above the gelation threshold leads to dramatically increased gel
rigidity, making the observation of fluctuation signals above the noise
difficult. After calibrating the spectra with the sensitivity factors
determined for each bead (see Fig.~5), the scatter in the spectra
typically remained large, unlike for actin. The inset in Fig.~13 shows
PSDs at 1~Hz for different bead sizes in 2.5\% gels, indicating this
large scatter.

\begin{figure}
\epsfxsize=\columnwidth
\centerline{\epsfbox{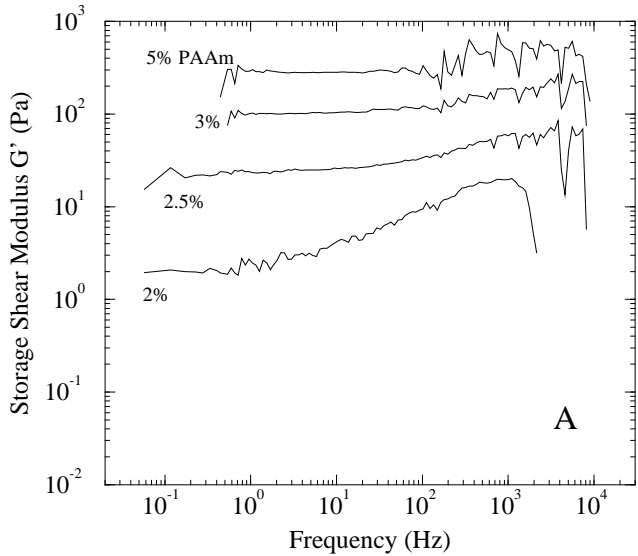}}
\epsfxsize=\columnwidth
\centerline{\epsfbox{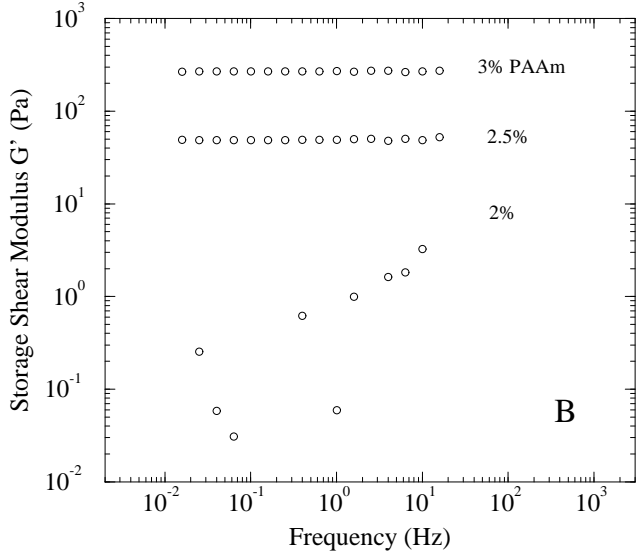}}
\caption[]{(A) Concentration dependence of the storage modulus in PAAm gels,
obtained from the spectra in Fig.~10. Moduli were calculated for
individual beads and linearly averaged. Elastic plateaus are clearly
visible and their extent is found to increase with gel concentration. The
vertical range is chosen to match (B).
(B) Control experiments with a cone-and-plate rheometer (Rheometrics RFS
II Fluids Spectrometer) with PAAm gels prepared in a manner identical to
the samples in (A). Forced oscillatory measurements were made at 1\%
strain and at 25$^\circ$C (data courtesy of M.~Osterfield, J.~Shah, and
P.~Janmey).}
\end{figure}

The 2 and 2.5\% gels are not much above the gelation threshold and the
gel rigidity there changes very rapidly with concentration. Therefore,
the observed scatter is presumably partly due to local gel
inhomogeneities. Bead polydispersity, as discussed in Materials and
Experimental Methods, is a further contribution. As for actin, we find
no correlation with distance from the substrate surface. The calculated
frequency-dependent storage moduli for the data in Fig.~10 are shown in
Fig.~11A. The transformation Eq.~(7) has largely eliminated unphysical
low-frequency noise in these spectra, allowing the plateau modulus
$G'(\omega\rightarrow0)$ to be easily estimated. For 2, 2.5, 3, and 5\%
gels the plateau values of $G'$ are approximately 2.0, 24, 100, and
280~Pa.

The effective scaling of $G'$ with concentration is 1.8 (for
concentrations between 3 and 5\%), and steeper for lower concentrations.
Our 2.5\% gels thus lie in a regime of very rapidly increasing stiffness
indeed (this is consistent with Ref. 49). For larger plateau moduli the
upturn shifts, as expected, to higher frequencies. The pure scaling
regime beyond the plateau is not reached for any of the samples,
although its existence is suggested by the slope of $-1.5$ in the PSDs.
The corresponding scaling of $G'(\omega)\propto\omega^{1/2}$ is
consistent with the Rouse model$^{4}$.

Fig.~11B shows the results of control experiments performed
with conventional cone-and-plate rheology (courtesy of M.~Osterfield,~
J.~Shah, P.~Janmey). Samples were prepared following the same recipes as
for the microrheological experiments. Plateau values agree within a
factor of 2-3. The data show qualitatively the same steep dependence on
concentration as the microrheological results. Elastic plateaus are
clearly visible for 2.5 and 3\% gels whereas for the 2\% gels
instrumental limits are reached, as seen from the large scatter of the
data.

\begin{figure}
\epsfxsize=\columnwidth
\centerline{\epsfbox{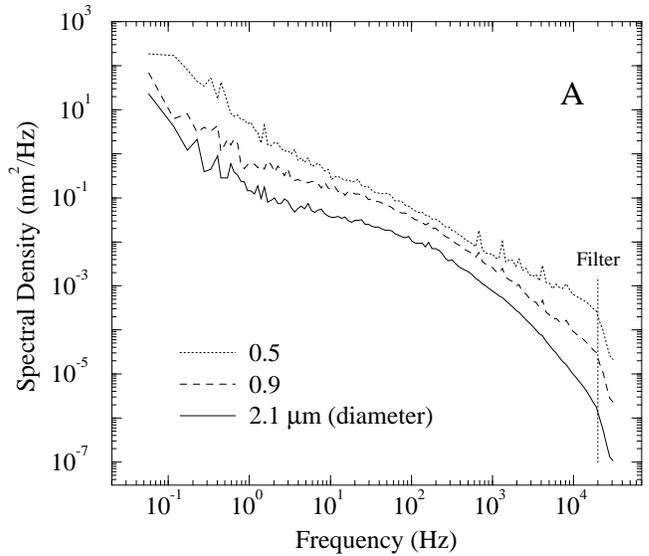}}
\epsfxsize=\columnwidth
\centerline{\epsfbox{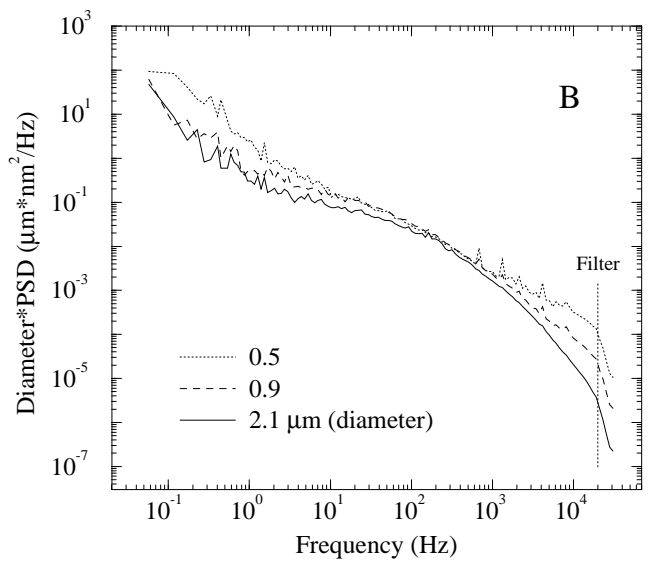}}
\caption[]{(A) Bead-size dependence of PSDs, for beads in PAAm gels of 2.5\%
concentration. Spectra for 0.5, 0.9 and 2,1~$\mu$m bead diameters are
averaged over 6, 7 and 10 beads respectively.  (B) Averaged PSDs from
(A) multiplied by their respective bead diameters.}
\end{figure}

Fig.~12A shows averaged spectra for different bead sizes
(between 0.5 and 2.1~$\mu$m) in 2.5\% PAAm gels; in
Fig.~12B these power spectra are multiplied by the
respective bead diameters. The $R^{-1}$-dependence in
Eq.~(5) predicts that the rescaled curves have the same
shape and magnitude. This is approximately true for the central
frequencies in the spectra, but the high rigidity of the PAAm gels
causes non-displacement noise to be significant both at high frequencies
(electronic detection noise) and low frequencies (thermal drifts, laser
beam-pointing and mode instabilities, etc.). The parts of the spectra
that are dominated by noise show no scaling. Due to the bead-size
dependence of the detector sensitivity, the low-frequency noise extends
higher for smaller beads.

The storage moduli $G'(\omega)$ for three different bead sizes 2.5\%
gels are shown in Fig.~13(inset). In contrast to what was found for actin, the
elastic plateau is clearly visible for all samples, and agrees for
different bead sizes in the same concentration to within about a factor
of 2. The elastic plateau extends to about the same frequency for all
bead sizes. This is in contrast to the low-frequency slope changes in
actin, which we interpret as finite-mesh-size effects, supported by a
consistent bead-size dependence. With mesh sizes as small as in PAAm,
draining effects should not play a role, down to the lowest frequencies
we measure. However, the trend in the data is to show apparently smaller
moduli for smaller beads. We attribute this artifact to a remaining
sensitivity to noise in the integration procedure used to calculate
$G'$. The detector sensitivity decreases dramatically for smaller beads,
thus including more noise in the integral, leading to the underestimate
of $G'$. The plateau moduli estimated by the value of $G'$ at 1~Hz
(Fig.~13) are 36~Pa (2.1~$\mu$m), 24~Pa (0.9~$\mu$m), and 17~Pa
(0.5~$\mu$m bead diameter).

\begin{figure}
\epsfxsize=\columnwidth
\centerline{\epsfbox{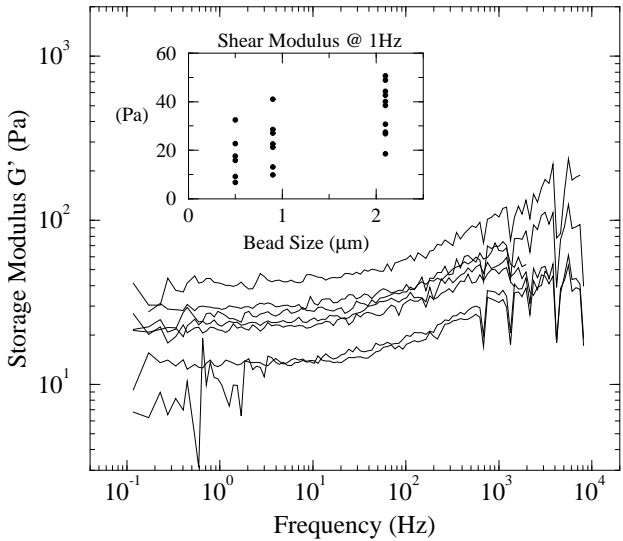}}
\caption[]{Variation in the storage modulus of 2.5\% PAAm gels, for seven
different 0.9~$\mu$m beads. The substantial scatter may be partly due to
bead polydispersity. Inset: Values of $G'$ at 1~Hz for multiple beads of
three sizes.}
\end{figure}

\subsection*{Discussion}

We have introduced the direct use of the Kramers-Kronig dispersion
integral Eq.~(7) to obtain storage and loss moduli from power spectra of
thermal motions. Other methods have been used to obtain shear moduli
from the thermal mean-square particle displacement as a function of
time: Mason and Weitz give an approximate method based on real-valued
Laplace transforms$^{28}$; Mason and Wirtz describe a somewhat different
approximation$^{29,50}$. A limitation of the former method is that a
specific functional form was postulated, fitted to a numerical Laplace
transform of particle displacement data, and then analytically continued
to the Fourier domain. Possible systematic errors in such a scheme are
not known. Such transformation methods have previously been shown to be
sensitive to noise and to behavior at the frequency extremes$^{51}$. In
contrast with Ref.~28, we have applied the transforms described in
Eqs.~(5) and (8) directly to our measured spectra. Both of these
transform methods are sensitive to the frequency extremes, but we
believe that the direct transformations we have employed are better
controlled.

In order to demonstrate the dependence of our Kramers-Kronig
transformation on the limits of the integration interval, we show in
Fig.~14 the transformation of a model power spectrum that
decreases as a power of frequency, above a corner frequency that we
arbitrarily place at 0.1~Hz.  We choose the power law to be
$\omega^{-1.75}$, similar to the power law observed in our fluctuation
power spectra for actin solutions.  This model spectrum corresponds,
theoretically, to a complex modulus $G^*$ that increases as $G^* \propto
\omega^{0.75}$ above the corner frequency, and a plateau in the storage
modulus $G'$ below the corner frequency.  We show in Fig.~14
the results of a transformation carried out with various values of the
upper and lower cutoffs.  The effect of cutoffs is minimal in the
central portion of the transforms.

\begin{figure}
\epsfxsize=\columnwidth
\centerline{\epsfbox{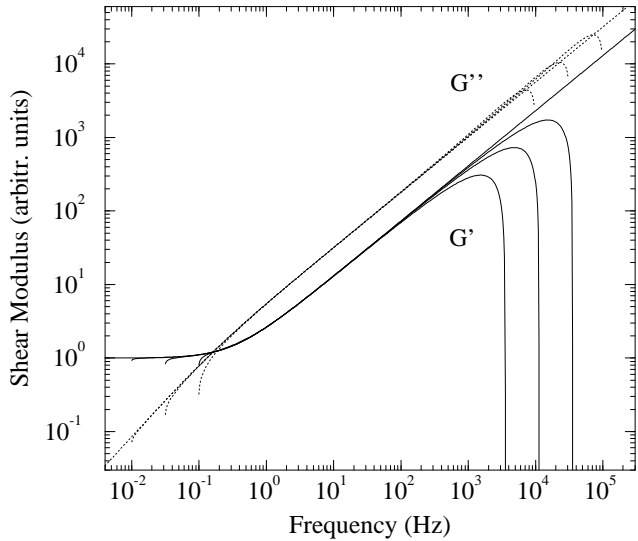}}
\caption[]{Frequency-cutoff dependence of the Kramers-Kronig transformation on a
model spectrum. The model PSD (not shown) decreases as $\omega^{-1.75}$
above a corner frequency at 0.1~Hz. The same Kramers-Kronig
transformation and complex inversion are performed as on the actual data
in Fig.~9, resulting in the storage and loss moduli $G'$ and $G''$. The
lower cutoffs are at $10^{-2}$, $10^{-1.5}$, and $10^{-1}$~Hz, and the
upper cutoffs are at $10^{4}$, $10^{4.5}$, and $10^{5}$~Hz. For
comparison, the curves without any cutoff are also included. Deviations
from the correct curve extend up by about a factor of 2 from the lower
cutoff and down by about a factor of 10 from the upper cutoff.  Our
actual data in Fig.~9 have an input frequency range similar to the curve
cut off at $10^{-1.5}$~Hz and $10^{4.5}$~Hz.}
\end{figure}

The shear elastic storage and loss moduli ($G'$ and $G''$) we found for
actin solutions are consistent with other experiments$^{26,27}$ in the
frequency regime where they can be compared. Other reported techniques
have been limited in frequency to a maximum of about 10~Hz.
Discrepancies persist in the literature,$^{16}$ and values up to 300~Pa
have been measured for F-actin at a concentration of 2~mg/ml. We did not
observe changes in viscoelastic behavior in F-actin over periods of at
least 24 hours after polymerization. This suggests that the local
properties of the solutions probed by the micron-sized beads are not
affected by slow network changes. On the other hand, macroscopic
techniques which have observed such changes may not have probed the
linear response regime.

Actin microrheology by others$^{26}$ could be interpreted as consistent
with $\omega^{3/4}$ scaling, although the authors there suggest
$\omega^{1/2}$ scaling; the frequencies observed were close to the onset
of the plateau. Macrorheological work$^{15,19}$ has also reported
$G^*\sim\omega^{1/2}$. Here too, the frequencies studied may have been
in the transition between plateau and scaling regime. New results from
multiple light scattering experiments (D. Weitz, personal communication)
are consistent with a scaling exponent close to $3/4$ at even higher
frequencies than we are able to measure. This scaling behavior in
semiflexible actin solutions may be a universal property of this type of
network, but is so far unexplained. In particular, there is no reason to
expect Rouse or Zimm scaling in these systems$^{4}$.  Bead dynamics in
actin networks consistent with the power spectra we observe have also
been reported in the limit of very dilute gels (specifically, for
$R\simeq\xi$)$^{25}$; the authors suggest a model based on single
filament dynamics. We believe that at the (higher) concentrations we
used, for which the mesh size is substantially smaller than bead
diameters, our continuum elastic approach is correct.

For solutions of semiflexible polymers, where effective entanglement
lengths may be much larger than the mesh size,$^{52,53}$ crosslinking is
expected to have a very strong effect on the gel rigidity. In cells, the
actin cortex is extensively crosslinked while the average filament
length is short. Therefore, a rigidity relevant for cell biology has to
be measured with a crosslinked actin gel, potentially lessening the
discrepancies between reported measurements.

In both F-actin solutions and PAAm gels, one might expect to be able to
resolve local inhomogeneities in viscoelastic parameters with
increasingly smaller beads. We did see considerable scatter in the
fluctuation power spectra, but this spread did not increase for smaller
beads, as one might have expected. Therefore the inhomogeneities in the
systems we studied may be smaller than effects of bead polydispersity
and other instrumental errors. More extensive statistical
characterization of the probe particles and improvement of the
instrument will reduce the noise, as well as increase the available
frequency range.  These impovements will allow closer examination of
sample inhomogeneities.

Our PAAm gels were very close to the gelation threshold and therefore
expected to be inhomogeneous.  For gels of 2, 2.5, 3, and 5\% we indeed
find a very steep increase in $G'$ (Fig.~15). This qualitative behavior
and the absolute values are consistent with our control experiments with
cone-and-plate rheology and the data match literature values in the
3-5\% range$^{49}$.  Macroscopic methods tend to get inaccurate for low
shear moduli on the order of 1~Pa, whereas our microscopic method
underestimates shear moduli above 100~Pa (stiff gels) due to noise in
the bead position detection.

\begin{figure}
\epsfxsize=\columnwidth
\centerline{\epsfbox{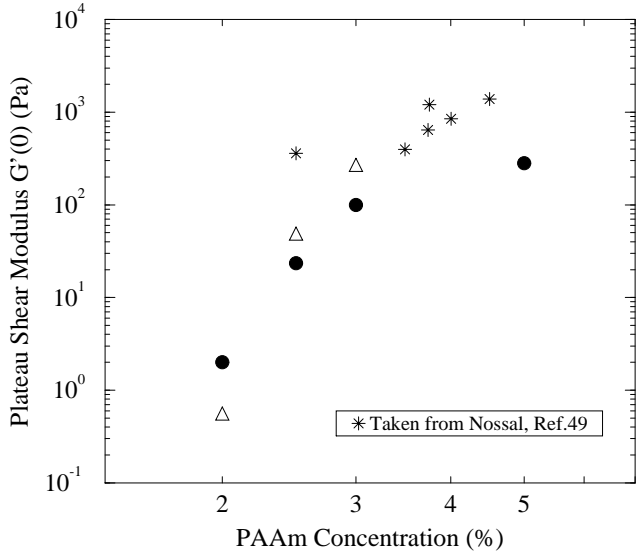}}
\caption[]{Storage shear modulus of PAAm as a function of gel concentration.
Solid circles: Microrheology results, values estimated from the plateaus
in Fig.~11A.  Open triangles: Cone-and-plate rheology (data courtesy of
M.~Osterfield, J.~Shah, and P.~Janmey), values estimated from the
plateaus in Fig.~11B.  Stars: Values taken from Ref.~[49], obtained from
dynamic light scattering, with a crosslinker fraction of 0.026, similar 
to our 3\% preparation.}
\end{figure}

In PAAm, the scaling of $G^*(\omega)\propto\omega^{1/2}$ is reflected
by an apparent power law slope of $-1.5$ in the spectra, consistent
with the Rouse model$^{4}$.

A current limitation of our method, so far unexplained, is a steepening
of the power spectrum that we consistently observe above a few kHz. We
believe this is an artifact, because it occurs even for beads in pure
water. We have ruled out electronic filtering as a cause. This effect
translates, in the Kramers-Kronig transformation of Eq.~(7), to a sharp
plunge in the calculated storage modulus at high frequency, which is of
course unphysical. Again, this illustrates the subtle nature of the
frequency sensitivity of the Kramers-Kronig and related transformations.
Although the downturn in our spectra occurs in the vicinity of 10~kHz,
its effects are apparent in $G'(\omega)$ approximately one decade below
this.

\subsection*{Conclusions}

We have used thermal fluctuations of micron-sized probes to measure
local properties of polymer solutions, by a passive and thus
non-invasive method, particularly appropriate for labile biopolymers.
Laser interferometry provides dynamic measurements of probe
displacements with high resolution and up to frequencies not previously
explored for F-actin solutions. With probes that are large compared to
the mesh size of the network, macroscopic properties such as the complex
shear modulus $G^*(\omega)$ can be estimated from the power spectrum of
fluctuations by using dispersion relations from linear response theory.
This analysis is made possible by our large detection bandwidth.  By
sampling micron-sized volumes we can in principle map out spatial
inhomogeneities in soft polymer systems.

Probes of sizes on the order of the mesh can be used to observe
deviations from macroscopic behavior, for example the transition to
microscopic filament dynamics. Observing this breakdown of continuum
elasticity may be of fundamental importance for understanding the
origins of elasticity in polymer systems, and has been the subject of
very recent theoretical attention$^{54-56}$. Furthermore, we have shown
that our technique (like other microrheological methods using probe
particles in polymer solutions) is predicted to be limited at low
frequencies---at least in measuring {\it macroscopic} shear moduli. This
is because of expected draining dynamics of network-plus-polymer, which
is not apparent in macroscopic rheometric methods.  On the other hand,
inertial effects of both probe and solvent, typically a problem in
sensitive macroscopic instruments,$^{1}$ occur only at MHz frequencies
in microscopic experiments.

On the micron length scale, a bead in a network should behave somewhat
like an organelle suspended in the cytoskeleton of a biological cell,
and we expect that the method will be transferable to measurements in
living cells, where biological processes are expected to strongly modify
viscoelastic parameters. Our technique is currently limited by broadband
(mainly acoustic) noise to rather soft materials, up to shear moduli of
a few hundred Pascal. This limitation will be extended by technical
improvements. Meanwhile, much remains to be done in developing a
theoretical understanding of the semiflexible polymer systems.

\subsection*{Acknowledgements}

This work was supported in part by the Whitaker Foundation, the National
Science Foundation (Grant Nos.\ BIR 95-12699 and DMR 92-57544), and by
the donors of the Petroleum Research Fund, administered by the ACS. We
thank M.~Osterfield, J.~Shah, and P.~Janmey for providing macroscopic
rheological data for controls, and P. Janmey for a control batch of
actin. We acknowledge generous technical support from the Rowland
Institute for Science, particularly by W. Hill. We thank G. Tank and J.
Langmore for help with the electron microscopy, and P. Olmsted, P.
Janmey, J. K\"as, A.C. Maggs, and D. Weitz for helpful discussions. FCM
also wishes to thank the Aspen Center for Physics.




\begin{thebibliography}{99}
\bibitem{31} The theoretical aspects of the microrheological technique described
here will be treated further in a separate publication$^{34}$.
\end{thebibliography}
\end{document}